\documentclass[12pt,a4paper,openany]{article}

\usepackage{graphicx}

\begin{document}
\title{The Astrocyte as a Gatekeeper of Synaptic Information Transfer}

\author{Vladislav Volman$^{1}$, Eshel Ben-Jacob$^{1,2}$ \& Herbert
Levine$^2$
\newline
\\
\centerline{1 - School of Physics and Astronomy, Raymond and
Beverly Sackler Faculty of Exact Sciences,}
\\
\centerline{Tel-Aviv Univ.,69978, Tel-Aviv, Israel}
\\
\centerline{2 - Center for Theoretical Biological Physics,
University of California at San Diego,}
\\
\centerline{La Jolla, CA 92093-0319 USA}
\\
\centerline{e-mails: volman(at)salk.edu, eshel(at)tamar.tau.ac.il,
hlevine(at)ucsd.edu}}
%

\baselineskip 18pt

\bibliographystyle{unsrt}

\maketitle
\section*{{\Large Abstract}}
We present a simple biophysical model for the coupling between
synaptic transmission and the local calcium concentration on an
enveloping astrocytic domain. This interaction enables the
astrocyte to modulate the information flow from presynaptic to
postsynaptic cells in a manner dependent on previous activity at
this and other nearby synapses. Our model suggests a novel,
testable hypothesis for the spike timing statistics measured for
rapidly-firing cells in culture experiments.

\newpage

\section*{{\Large Introduction}}
In recent years, evidence has been mounting regarding the possible
role of glial cells in the dynamics of neural tissue
~\cite{Volterra05,Haydon01,Newman03,RossiTakano06}. For astrocytes
in particular, the specific association of processes with synapses
and the discovery of two-way astrocyte-neuron communication has
demonstrated the inadequacy of the previously-held view regarding
the purely supportive role for these glial cells. Instead, future
progress requires rethinking how the dynamics of the coupled
neuron-glial network can store, recall, and process information.

At the level of cell biophysics, some of the mechanisms underlying
the so-called "tripartite synapse"~\cite{Araque99} are becoming
more clear. For example, it is now well-established that
astrocytic mGlu receptors detect synaptic activity and respond via
activation of the calcium-induced calcium release pathway, leading
to elevated $Ca^{2+}$ levels. The spread of these levels within a
micro-domain of one cell can coordinate the activity of disparate
synapses that are associated with the same
micro-domain~\cite{Perea02}. Moreover, it might even be possible
to transmit information directly from domain to domain and even
from astrocyte to astrocyte, if the excitation level is strong
enough to induce either intracellular or intercellular calcium
waves~\cite{Cornell-Bell90,Charles91,Cornell-Bell91}. One sign of
the maturity in our understanding is the formulation of
semi-quantitative models for this aspect of neuron-glial
communication~\cite{Jung04,Sneyd95,Hofer02}.

There is also information flow in the opposite direction, from
astrocyte to synapse. Direct experimental evidence for this, via
the detection of the modulation of synaptic transmission as a
function of the state of the glial cells will be reviewed in more
detail below. One of the major goals of this work will be to
introduce a simple phenomenological model for this interaction.
The model will take into account both a deterministic effect of
high $Ca^{2+}$ in the astrocytic process, namely the reduction of
the post-synaptic response to incoming spikes on the presynaptic
axon~\cite{Araqueejn98}, and a stochastic effect, namely the
increase in the frequency of observed miniature post-synaptic
current events uncorrelated with any input~\cite{Araquejns98}.
There are also direct NMDA-dependent effects on the postsynaptic
neuron of astrocyte-emitted factors~\cite{Perea05}, which are not
considered here.

As we will show, the aforementioned coupling allows the astrocyte
to act as a "gatekeeper" for the synapse. By this, we mean that
the amount of data transmitted across the synapse can be modulated
by astrocytic dynamics. These dynamics may be controlled mostly by
other synapses, in which case the gate-keeping will depend on
dynamics external to the specific synapse under consideration.
Alternatively, the dynamics may depend mostly on excitation from
the selfsame synapse, in which case the behavior of the entire
system is determined self-consistently. Here we will focus on the
latter possibility and leave for future work the discussion of how
this mechanism could lead to multi-synaptic coupling

Our ideas regarding the role of the astrocyte are utilized to
offer a new explanation for observations regarding firing patterns
in cultured neuronal networks. In particular, spontaneous bursting
activity in these networks is regulated by a set of rapidly firing
neurons which we refer to as "spikers"; these neurons exhibit
spiking even during long inter-burst intervals and hence must have
some form of self-consistent self-excitation. We model these
neurons as containing astrocyte-mediated self-synapses (autapses)
(see~\cite{Segal91,Segal94,Bekkers91}) and show that this
hypothesis naturally accounts for the observed unusual inter-spike
interval distribution. Additional tests of this hypothesis will be
proposed at the end.

\section*{{\Large Experimental observations}}
{\bf Cultured neuronal networks:} The cultured neuronal networks
presented here are self-generated from dissociated cultures of
mixed cortical neurons and glial cells drawn from one-day-old
Charles River rats. The dissection, cell dissociation and
recording procedures were previously described in
detail~\cite{Levy2002}. Briefly, following dissection, neurons are
dispersed by enzymatic treatment and mechanical dissociation. Then
the cells are homogeneously plated on multi electrode arrays (MEA,
Multi-Channel Systems), pre-coated with Poly-L-Lysine. Culture
media was DMEM, (sigma) enriched by serum and changed every two
days. Plated cultures are placed on the MEA board (B-MEA-1060,
Multi Channel Systems) for simultaneous long-term noninvasive
recordings of neuronal activity from several neurons at a time.
Recorded signals are digitized and stored for off-line analysis on
a PC via an A-D board (Microstar DAP) and data acquisition
software (Alpha-Map, Alpha Omega Engineering).  Non-invasive
recording of the networks activity (action potentials) is possible
due to the capacitive coupling that some of the neurons form with
some of the electrodes. Since typically one electrode can record
signals from several neurons, a specially developed spike-sorting
algorithm \cite{Hulata00} is utilized to reconstruct {\em single}
neuron-specific spike series. Although there are no externally
provided guiding stimulations or chemical cues, relatively intense
dynamical activity is spontaneously generated within several days.
The activity is marked by the formation of synchronized bursting
events (SBEs) Ð short (~200ms) time windows during which most of
the recorded neurons participate in relatively rapid
firing~\cite{Segev01}. These SBEs are separated by long intervals
(several seconds or more) of sporadic neuronal firing of most of
the neurons. A few neurons (referred to as Òspiker neuronsÓ)
exhibit rapid firing even during the inter SBEs time intervals.
These neurons also exhibit much faster firing rates during the
SBEs and their inter-spike-intervals distribution is marked by a
long tail behavior (see Fig. \ref{fig-Fig4}).

{\bf Inter-spike interval (ISI) increments distribution:} One of
the tools used to compare model results with measured spike data
concerns the distribution of increments in the spike times,
defined as $\delta(i)=ISI(i+1)-ISI(i), i\geq 1$. The distribution
of $\delta(i)$ will have heavy tails if there are a wide range of
inter-spike intervals and if there are rapid transitions from one
type of interval to the next. For example, rapid transitions from
bursting events to occasional inter-burst firings will lead to
such a tail. Applying this analysis to the recorded spike data of
cultured cortical networks, Segev et al. \cite{Levy2002} found
that distributions of neurons ISI increments can be well-fitted
with Levy functions over 3 decades in time.

\section*{{\Large The model}}

In this section we present the mathematical details of the models
which will be employed in this work. Readers interested mainly in
the conclusions can skip directly to the Results section.

The basic notion we use is that standard synapse models must be
modified to account for the astrocytic modulation, depending of
course on the calcium level. In turn, the astrocytic calcium level
is affected by synaptic activity; for this we use the Li-Rinzel
model where the IP$_3$ concentration parameter governing the
excitability is increased upon neurotransmitter release. These
ingredients suffice to demonstrate what we mean by gate-keeping.
Finally, we apply this model to the case of an autaptic
oscillator, which requires the introduction of neuronal dynamics.
For this, we chose the Morris-Lecar model as a generic example of
a type-I firing system. None of our results would be altered with
a different choice, as long as we retain the tangent-bifurcation
structure which allows for arbitrarily long inter-spike intervals.
Now for the details:

{\bf TUM Synapse Model:} To describe the kinetics of a synaptic
terminal, we have used the model of an activity-dependent synapse
first introduced by Tsodyks, Uziel and Markram \cite{Tsodyks00}.
In this model, the effective synaptic strength evolves according
to the following equations :
\begin{eqnarray}\label{eq:Tsodyks_X_Eq}
\dot{x} &= & \frac{z}{\tau_{rec}}-ux\delta(t-t_{sp}) \nonumber \\
\dot{y} & = & -\frac{y}{\tau_{in}}+ux\delta(t-t_{sp})\nonumber \\
\dot{z} &= & \frac{y}{\tau_{in}}-\frac{z}{\tau_{rec}}
\end{eqnarray}
Here, $x$, $y$, and $z$ are the fractions of synaptic resources in
the recovered, active and inactive states, respectively. For an
excitatory glutamatergic synapse, the values attained by these
variables can be associated with the dynamics of vesicular
glutamate. As an example, the value of $y$ in this formulation
will be proportional to the amount of glutamate that is being
released during the synaptic event, and the value of $x$ will be
proportional to the size of readily releasable vesicle pool. The
time-series $t_{sp}$ denote the arrival times of pre-synaptic
spikes, $\tau_{in}$ is the characteristic time of post-synaptic
currents (PSCs) decay, and $\tau_{rec}$ is the recovery time from
synaptic depression. Upon arrival of a spike to the pre-synaptic
terminal at time $t_{sp}$, a fraction $u$ of available synaptic
resources is transferred from the recovered state to the active
state. Once in the active state, synaptic resource rapidly decays
to the inactive state, from which it recovers within a time-scale
$\tau_{rec}$. Since the typical times are assumed to satisfy
$\tau_{rec}>>\tau_{in}$, the model predicts onset of short-term
synaptic depression after a period of high-frequency repetitive
firing. The onset of depression can be controlled by the variable
$u$, which describes the effective use of synaptic resources by
the incoming spike. In the original TUM model, the variable $u$ is
taken to be constant for the excitatory post-synaptic neuron; in
what follows we will set $u=0.1$. Other parameter choices for
these equations as well as for the rest of the model equations are
presented in the accompanying table.

To complete the specification, it is assumed that the resulting
post-synaptic current (PSC), arriving at the model neurons' soma
through the synapse depends linearly on the fraction of available
synaptic resources. Hence, a total synaptic current seen by a
neuron is $I_{syn}(t)=Ay(t)$, where $A$ stands for an absolute
synaptic strength. At this stage, we do not take into account the
long-term effects associated with the plasticity of neuronal
somata and take the parameter $A$ to be time-independent.

{\bf Astrocyte response:} Astrocytes adjacent to synaptic
terminals respond to the neuronal action potentials by binding
glutamate to their metabotropic glutamate
receptors~\cite{Porter96}. The activation of these receptors then
triggers the production of $IP_{3}$, which, consequently, serves
to modulate the intracellular concentration of calcium ions; the
effective rate of $IP_{3}$ production depends on the amount of
transmitter that has been released during the synaptic event.

We therefore assume that the production of intracellular $IP_{3}$
in the astrocyte is given by
\begin{equation}
\frac{d[IP_{3}]}{dt}=\frac{IP_{3}^{*}-IP_{3}}{\tau_{ip_{3}}}+r_{ip_{3}}y
\end{equation}
The above equation is similar to the formulation used by Nadkarni
and Jung \cite{Jung04}, with some important differences. First,
the effective rate of $IP_{3}$ production depends not on the
potential of neuronal membrane, but on the amount of
neurotransmitter that is being released into the synaptic cleft.
Hence, as the resources of synapse are depleted (due to
depression), there will be less transmitter released, and,
therefore, the $IP_{3}$ will be produced at lower rates, leading
eventually to decay of calcium concentration. Second, as the
neuro-transmitter is released also during spontaneous synaptic
events (noise), the latter will also influence the production of
$IP_{3}$ and subsequent calcium oscillations.

{\bf Astrocyte:} To model the dynamics of a single astrocytic
domain, we use the Li-Rinzel model \cite{LiRinzelModel,Jung04},
which has been specifically developed to take into account the
$IP_{3}$-dependent dynamical changes in the concentration of
cytosolic $Ca^{2+}$. This is based on the theoretical studies of
Nadkarni and Jung \cite{Jung04}, where it is decisively
demonstrated that astrocytic $Ca^{2+}$ oscillations may account
for the spontaneous activity of neurons.

The intracellular concentration of $Ca^{2+}$ in the astrocyte is
described by the following set of equations:
\begin{equation}
\frac{d[Ca^{2+}]}{dt}=-J_{chan}-J_{pump}-J_{leak}
\end{equation}

\begin{equation}
\frac{dq}{dt}=\alpha_{q}(1-q)-\beta_{q}q
\end{equation}
Here, $q$ is the fraction of activated $IP_{3}$ receptors. The
fluxes of currents through ER membrane are given in the following
expressions:
\begin{equation}
J_{chan}=c_{1}v_{1}m_{\infty}^{3}n_{\infty}^{3}q^{3}([Ca^{2+}]-[Ca^{2+}]_{ER})
\end{equation}
\begin{equation}
J_{pump}=\frac{v_{3}[Ca^{2+}]^{2}}{k_{3}^{2}+[Ca^{2+}]^{2}}
\end{equation}
\begin{equation}
J_{leak}=c_{1}v_{2}([Ca^{2+}]-[Ca^{2+}]_{ER})
\end{equation}
where
\begin{equation}
m_{\infty}=\frac{[IP_{3}]}{[IP_{3}]+d_{1}}
\end{equation}
\begin{equation}
n_{\infty}=\frac{[Ca^{2+}]}{[Ca^{2+}]+d_{5}}
\end{equation}
\begin{equation}
\alpha_{q}=a_{2}d_{2}\frac{[IP_{3}]+d_{1}}{[IP_{3}]+d_{3}}
\end{equation}
\begin{equation}
\beta_{q}=a_{2}[Ca^{2+}]
\end{equation}
The reversal $Ca^{2+}$ concentration ($[Ca^{2+}]_{ER}$) is
obtained after requiring conservation of the overall $Ca^{2+}$
concentration:
\begin{equation}
[Ca^{2+}]_{ER}=\frac{c_{0}-[Ca^{2+}]}{c_{1}}
\end{equation}

{\bf Glia-synapse interaction:} Astrocytes affect synaptic vesicle
release in a calcium dependent manner. Rather than attempt a
complete biophysical model of the complex chain of events leading
from calcium rise to vesicle release~\cite{Gandhi03}, we proceed
in a phenomenological manner. We define a dynamical variable $f$
which phenomenologically will capture this interaction; when the
concentration of calcium in its synapse-associated process exceeds
a threshold, we assume that the astrocyte emits a finite amount of
neurotransmitter into the peri-synaptic space, thus altering the
state of a nearby synapse; this interaction occurs via glutamate
binding to pre-synaptic mGlu and NMDA receptors \cite{Zhang04}. As
the internal astrocyte resource of neurotransmitter is finite, we
include saturation term $(1-f)$ in the dynamical equation for $f$.
The final form is
\begin{equation}\label{eq:CaStateVarEq}
\dot{f}=\frac{-f}{\tau_{Ca^{2+}}}+(1-f)\kappa\Theta([Ca^{2+}]-[Ca^{2+}_{threshold}])
\end{equation}

Given this assumption, equations \ref{eq:Tsodyks_X_Eq} should be
modified to take this modulation into account. We assume the
following simple form:
\begin{equation}
\dot{x}=\frac{z}{\tau_{rec}}-(1-f)ux\delta(t-t_{sp})-x\eta(f)
\end{equation}
\begin{equation}
\dot{y}=\frac{-y}{\tau_{in}}+(1-f)ux\delta(t-t_{sp})+x\eta(f)
\end{equation}
In the above equations, $\eta(f)$ represents a noise term
modelling the increased occurrence of mini-PSC's. The fact that a
noise increase accompanies an amplitude decrease is partially due
to competition for synaptic resources between this two release
modes~\cite{Otsu04}. Based on experimental observations, we
prescribe that the dependence of $\eta(f)$ on $f$ is such that
rate of noise occurrence (the frequency of $\eta(f)$ in a fixed
time step) increases with increasing $f$, but the amplitude
distribution (modelled here as a Gaussian-distributed variable
centered around positive mean) remains unchanged. For the rate of
noise occurrence, we chose the following functional dependence:
\begin{equation}
P(f)=P_{0}exp(-(\frac{1-f}{\sqrt{2}\sigma})^{2})
\end{equation}
with $P_{0}$ representing the maximal frequency of $\eta(f)$ in a
fixed time step.

Note that although both synaptic terminals and astrocytes utilize
glutamate for their signaling purposes, we assume the two
processes to be independent. In so doing, we rely on the existing
biophysical experiments which demonstrate that, whereas a
presynaptic terminal releases glutamate in the synaptic cleft,
astrocytes selectively target extra-synaptic glutamate
receptors~\cite{Araqueejn98,Araquejns98}. Hence, synaptic
transmission does not interfere with the astrocyte-to-synapse
signalling.

{\bf Neuron model:} We  describe the neuronal dynamics with a
simplified two-component Morris-Lecar model~\cite{MorrisLecar81} :
\begin{equation}\label{eq:ML_voltage}
\dot{V}=-I_{ion}(V,W)+I_{ext}(t)
\end{equation}
\begin{equation}\label{eq:ML_fraction}
\dot{W}(V)=\phi\frac{W_{\infty}(V)-W(V)}{\tau_{W}(V)}
\end{equation}
with $I_{ion}(V,W)$ representing the contribution of the internal
ionic $Ca^{2+}$, $K^{+}$ and leakage currents with their
corresponding channel conductivities $g_{Ca}$, $g_{K}$ and $g_{L}$
being constant :
\begin{eqnarray}\label{eq:ML_current}
I_{ion}(V,W)=g_{Ca}m_{\infty}(V)(V-V_{Ca})+\nonumber\\\space+g_{K}W(V)(V-V_{K})+g_{L}(V-V_{L})
\end{eqnarray}
$I_{ext}$ represents all the external current sources stimulating
the neuron, such as signals received through its synapses,
glia-derived currents, artificial stimulations as well as any
noise sources. In the absence of any such stimulation, the
fraction of open potassium channels, $W(V)$, relaxes towards its
limiting curve (nullcline) $W_{\infty}(V)$, which is described by
the sigmoid function :
\begin{equation}\label{eq:ML_winf}
W_{\infty}(V)=\frac{1}{2}(1+tanh(\frac{V-V_{3}}{V_{4}}))
\end{equation}
within a characteristic time scale given by :
\begin{equation}\label{eq:ML_tau}
\tau_{W}(V)=\frac{1}{cosh(\frac{V-V_{3}}{2V_{4}})}
\end{equation}
In contrast to this, it is assumed in the Morris-Lecar model that
calcium channels are activated immediately. Accordingly, the
fraction of open $Ca^{2+}$ channels obeys the following equation :
\begin{equation}\label{eq:ML_minf}
m_{\infty}(V)=\frac{1}{2}(1+tanh(\frac{V-V_{1}}{V_{2}}))
\end{equation}

For an isolated neuron, rendered with a single autapse, one has
$I_{ext}(t)=I_{syn}(t)+I_{base}$ where $I_{syn}(t)$ is the current
arriving through the self-synapse, and $I_{base}$ is some constant
background current.  In this work, we assume that $I_{base}$ is
such that, when acting alone, it causes a neuron to fire at very
low constant rate. Of course these two terms enter the equation
additively and the dynamics just depends on the total external
current. Nonetheless it is important to separate these terms as
only one of them enters through the synapse; it is only this term
that is modulated by astrocytic glutamate release and only this
term that would be changed by synaptic blockers. As we will
mention later, the baseline current may also be due to astrocytes,
albeit to a direct current directed into the neuronal soma. In
anticipation of a better future understanding of this term, we
consider it separately from the constant appearing in leak current
($g_L V_L$) although there is clearly some redundancy in the way
these two terms set the operating point of the neuron.

\section*{{\Large Results}}
\begin{figure}
\centerline{
\resizebox{0.55\textwidth}{!}{\includegraphics{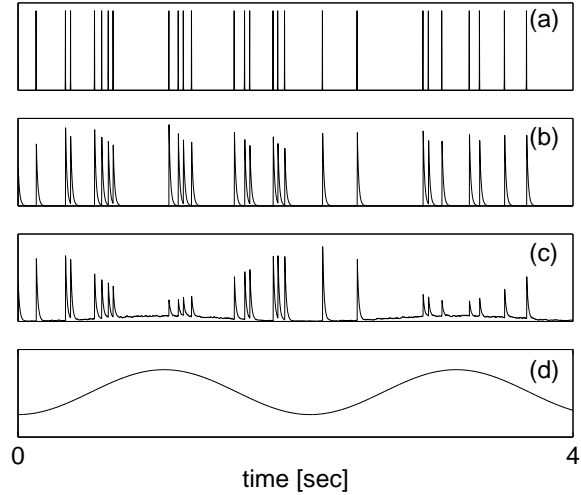}}}
\caption[]{ The generic effect of an astrocyte on the pre-synaptic
depression, as captured by our phenomenological model (see text
for details). To illustrate the effect of pre-synaptic depression
and the astrocyte influence, we feed a model synapse with the
input of spikes taken from the recorded activity of a cultured
neuronal network (see main text and \cite{Segev01} for details).
a) The input sequence of spikes that is fed into the model
pre-synaptic terminal. b) Each spike arriving at the model
pre-synaptic terminal results in the post-synaptic current
$(PSC)$. The strength of the post-synaptic current depends on the
amount of available synaptic resources, and the synaptic
depression effect is clearly observable during spike trains with
relatively high frequency. c) The effect of a periodic gating
function, $f(t)=0.5+f_{0}sin(wt)$, shown in (d). The period of the
oscillation, $T=\frac{2\pi}{\omega}=2 sec$, is taken to be
compatible with the typical time scales of variations in the
intra-glial $Ca^{2+}$ concentration. Note the reduction in the
$PSC$ near the maxima of $f$, along with the elevated base-line
resulting from the increase in the rate of spontaneous
pre-synaptic transfer.} \label{fig-Fig1}
\end{figure}

{\bf Synaptic Model:} In simple models of neural networks, the
synapse is considered to be a passive element which directly
transmits information, in the form of arriving spikes on the
pre-synaptic terminal, to post-synaptic currents. It has been
known for a long time that more complex synaptic dynamics can
affect this transfer. One such effect concerns the finite
reservoir of presynaptic vesicle resources and was modelled by
Tsodyks, Uziel and Markram (TUM)~\cite{Tsodyks00}. Spike trains
with too high a frequency will be attenuated by a TUM synapse, as
there is insufficient recovery from one arrival to the next. To
demonstrate this effect, we fed the TUM synaptic model with an
actual spike train recorded from a neuron in a cultured network
(shown in Fig. \ref{fig-Fig1}a); the resulting post-synaptic
current (PSC) is shown in Fig. \ref{fig-Fig1}b. As is expected
there is attenuation of the PSC height during time windows with
high rates of pre-synaptic spiking input.

{\bf The effect of pre-synaptic gating:}  Our goal is to extend
the TUM model to include the interaction of the synapse with an
astrocytic process imagined to be wrapped around the synaptic
cleft. The effects of astrocytes on stimulated synaptic
transmission are well-established. Araque {\em et
al.}~\cite{Araqueejn98} report that astrocyte stimulation reduced
the magnitude of action potential evoked excitatory and inhibitory
synaptic currents by decreasing the probability of evoked
transmitter release.  Specifically, pre-synaptic metabotropic
glutamate receptors (mGluRs) have been shown to affect the
stimulated synaptic transmission by regulating pre-synaptic
voltage-gated calcium channels which eventually leads to the
reduction of calcium flux during the incoming spike and results in
decrease of amplitude of synaptic transmission. These results are
best shown in Fig. 8 of their paper, which presents the amplitude
of evoked EPSC both before and after stimulation of an associated
astrocyte. Note that we are referring here to ``faithful"
synapses, i.e. synapses that transmit almost all of the incoming
spikes. Effects of astrocytic stimulation on highly stochastic
synapses, namely the increase in fidelity~\cite{Kang98}, are not
studied here.

In addition, astrocytes were shown to increase the frequency of
spontaneous synaptic events. In detail, Araque {\em et
al.}~\cite{Araquejns98} have shown that astrocyte stimulation
increases the frequency of miniature post-synaptic currents
(mPSC), without modifying their amplitude distribution, suggesting
that astrocytes act to increase the probability of vesicular
release from the pre-synaptic terminal. Although the exact
mechanism is unknown, this effect is believed to be mediated by
NMDA receptors located at the pre-synaptic terminal. It is
important to note that the two kinds of astrocytic influence on
the synapse (decrease of the probability of evoked release and
increase in the probability of spontaneous release) do not
contradict each other. Evoked transmitter release depends on the
calcium influx through calcium channels that can be inhibited by
the activation of pre-synaptic mGluRs. On the other hand, the
increase in the probability of spontaneous release follows because
of the activation of pre-synaptic NMDA channels. In addition,
spontaneous activity can deplete the vesicle pool (either in terms
of number or in terms of filling) and hence directly lower
triggered release amplitudes~\cite{Otsu04}.

We model these effects by two modifications of the TUM model.
First, we introduce a gating function $f$ which modulates the
stimulated release in a calcium-dependent manner. This term will
cause the synapse to turn off at high calcium. This pre-synaptic
gating effect is demonstrated in Fig. \ref{fig-Fig1}c, where we
show the resulting PSC corresponding to a case in which $f$ is
chosen to vary periodically with a time scale consistent with
astrocytic calcium dynamics. The effect on the recorded spike
train data is quite striking. The second effect, namely the
increase of stochastic release in the absence of any input, is
included as a $f$ dependent noise term in the TUM equations. This
will be important as we turn to a self-consistent calculation of
the synapse coupled to a dynamical astrocyte.

{\bf The gate-keeping effect:} We close the synapse-glia-synapse
feedback loop by inclusion of the effect of the pre-synaptic
activity on the intracellular $Ca^{2+}$ dynamics in the astrocyte
that in turn set the value of the gating function $f$. Nadkarni
and Jung~\cite{Jung04} have argued that the basic calcium
phenomenology in the astrocyte, arising via the glutamate-induced
production of $IP_3$, can be studied via the Li-Rinzel model. What
emerges from their work is that the dynamics of the
intra-astrocyte $Ca^{2+}$ level depends on the intensity of the
pre-synaptic spike train, acting as an information integrator over
a time scale on the order of seconds; the level of $Ca^{2+}$ in
the astrocyte increases according to the summation of the synaptic
spikes over time. If the total  number of spikes is low, the
$Ca^{2+}$ concentration in the astrocyte remains below a
self-amplification threshold level and simply decays back to its
resting level with some characteristic time. However, things
change dramatically when a sufficiently intense set of signals
arises across the synapse. Now, the $Ca^{2+}$ concentration
overshoots beyond its linear response level, followed by decaying
oscillations.

\begin{figure}
\centerline{
\resizebox{0.55\textwidth}{!}{\includegraphics{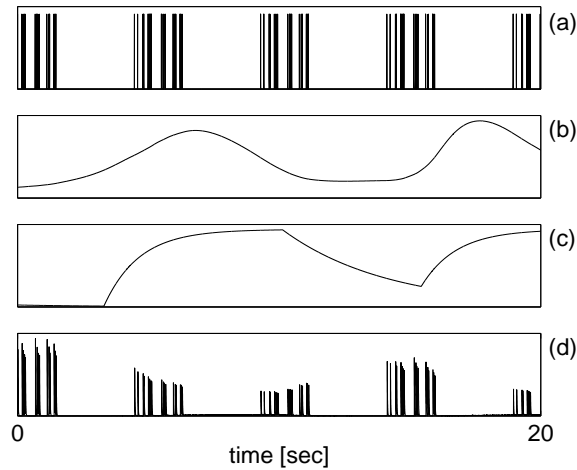}}}
\caption[]{ The "gate-keeping" effect in a glia-gated synapse. Top
panel (a) shows the input sequence of spikes, which is composed of
several copies of the sequence shown in figure \ref{fig-Fig1},
separated by segments of long quiescent time. The resulting time
series may be viewed as bursts of action potentials arriving at
the model pre-synaptic terminal. The first burst of spikes results
in the elevation of free astrocyte $Ca^{2+}$ concentration (shown
in (b)), but this elevation alone is not sufficient to evoke
oscillatory response. An additional elevation of $Ca^{2+}$,
leading to the emergence of oscillation, is provided by the second
burst of spikes arriving at the pre-synaptic terminal. Once the
astrocytic $Ca^{2+}$ crosses a pre-defined threshold, it starts to
exert a modulatory influence back on the pre-synaptic terminal. In
the model, this is manifested by the rising dynamics of the gating
function (shown in (c)). Note that, as the decay time of the
gating function $f$ is of the order of seconds, the astrocyte
influence on the pre-synaptic terminal persists even after
concentration of astrocyte $Ca^{2+}$ has fallen. This is best seen
from figure (d), where we show the profile of the post-synaptic
current $(PSC)$. The third burst of spikes arriving at the
pre-synaptic terminal is modulated due to the astrocyte, even
though the concentration of $Ca^{2+}$ is relatively low at that
time. This modulation extends also to the fourth burst of spikes,
which together with the third burst leads again to the oscillatory
response of astrocyte $Ca^{2+}$. Taken together, all of these
results illustrate a temporally non-local "gate-keeping" effect of
glia cells.} \label{fig-Fig2}
\end{figure}

Given our aforementioned results, these high $Ca^{2+}$ levels in
the astrocyte will in fact attenuate spike information that
arrives subsequent to strong bursts of activity. We illustrate
this time-delayed gate-keeping (TDGK) effect in Fig.
\ref{fig-Fig2}. We constructed a spike train by placing a time
delay in between segments of recorded sequences. As can be seen,
since the degree of activity during the first two segments exceeds
the threshold level, there is attenuation of the late-arriving
segments. Thus, the information passed through the synapse is
modulated by previous arriving data.

{\bf Autaptic Excitatory Neurons:} Our new view of synaptic
dynamics will have broad consequences for making sense of neural
circuitry. To illustrate this prospect, we turn to the study of an
autaptic oscillator~\cite{Seung00}, by which we mean an excitatory
neuron that exhibits repeated spiking driven at least in part by
self-synapses~\cite{Segal91,Segal94,Bekkers91,Lubke96}. By
including the coupling of a model neuron to our synapse system, we
can investigate both the case of the role of an associated
astrocyte with externally imposed temporal behavior and the case
where the astrocyte dynamics is itself determined by feedback from
this particular synapse. Finally, we should be clear that when we
refer to one synapse, we are also dealing implicitly with the case
of multiple self-synapses all of what are coupled to the same
astrocytic domain which in turn is exhibiting correlated dynamics
in its processes connecting to these multiple sites. It is
important to note that this same modulation can in fact correlate
multiple synapses connecting distinct neurons which are coupled to
the same astrocyte. The effect of this new multi-synaptic coupling
on the spatio-temporal flow of information in a model network will
be described elsewhere.

\begin{figure}
\centerline{
\resizebox{0.45\textwidth}{!}{\includegraphics{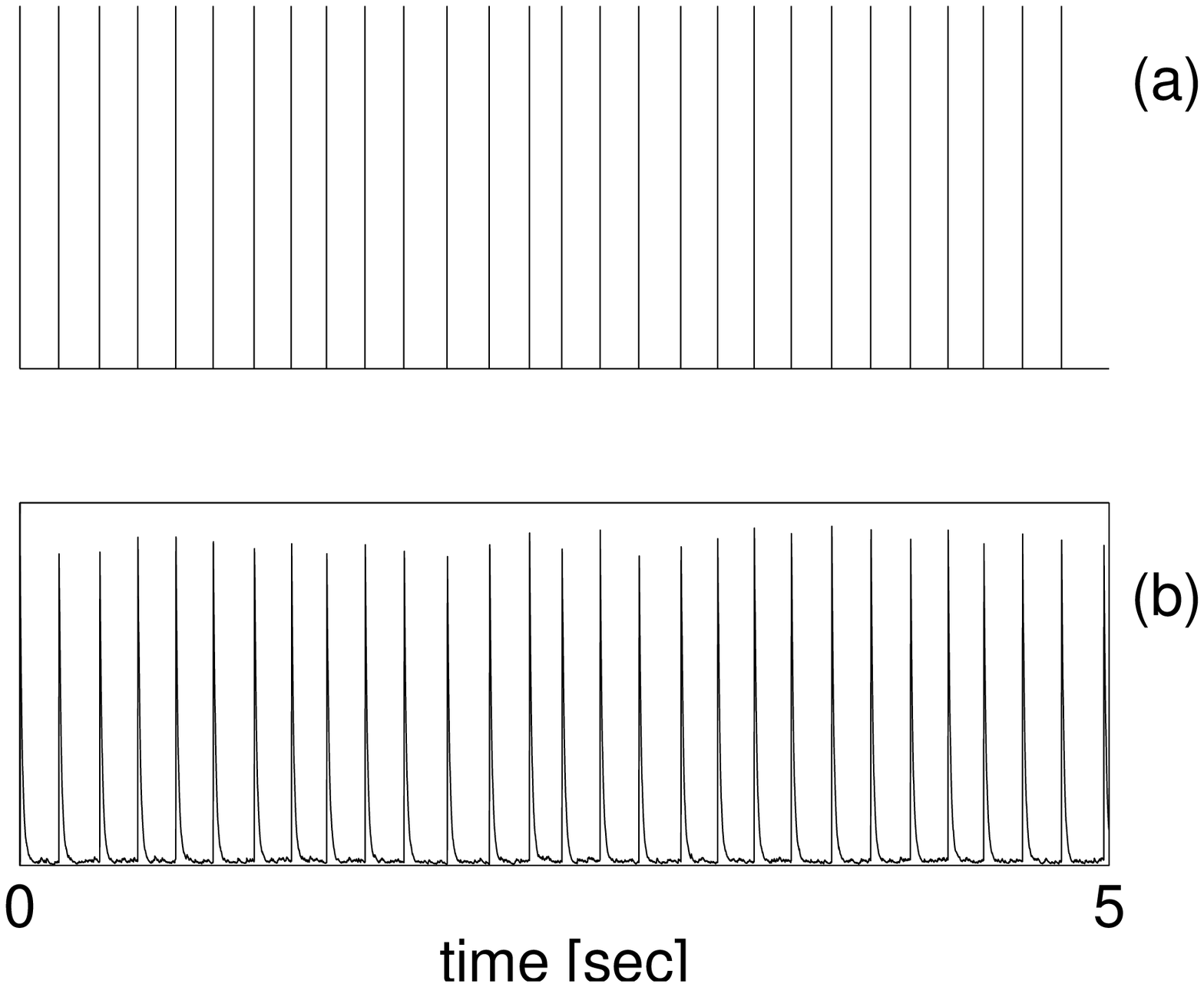}}}
\vspace{0.5cm}
\centerline{\resizebox{0.45\textwidth}{!}{\includegraphics{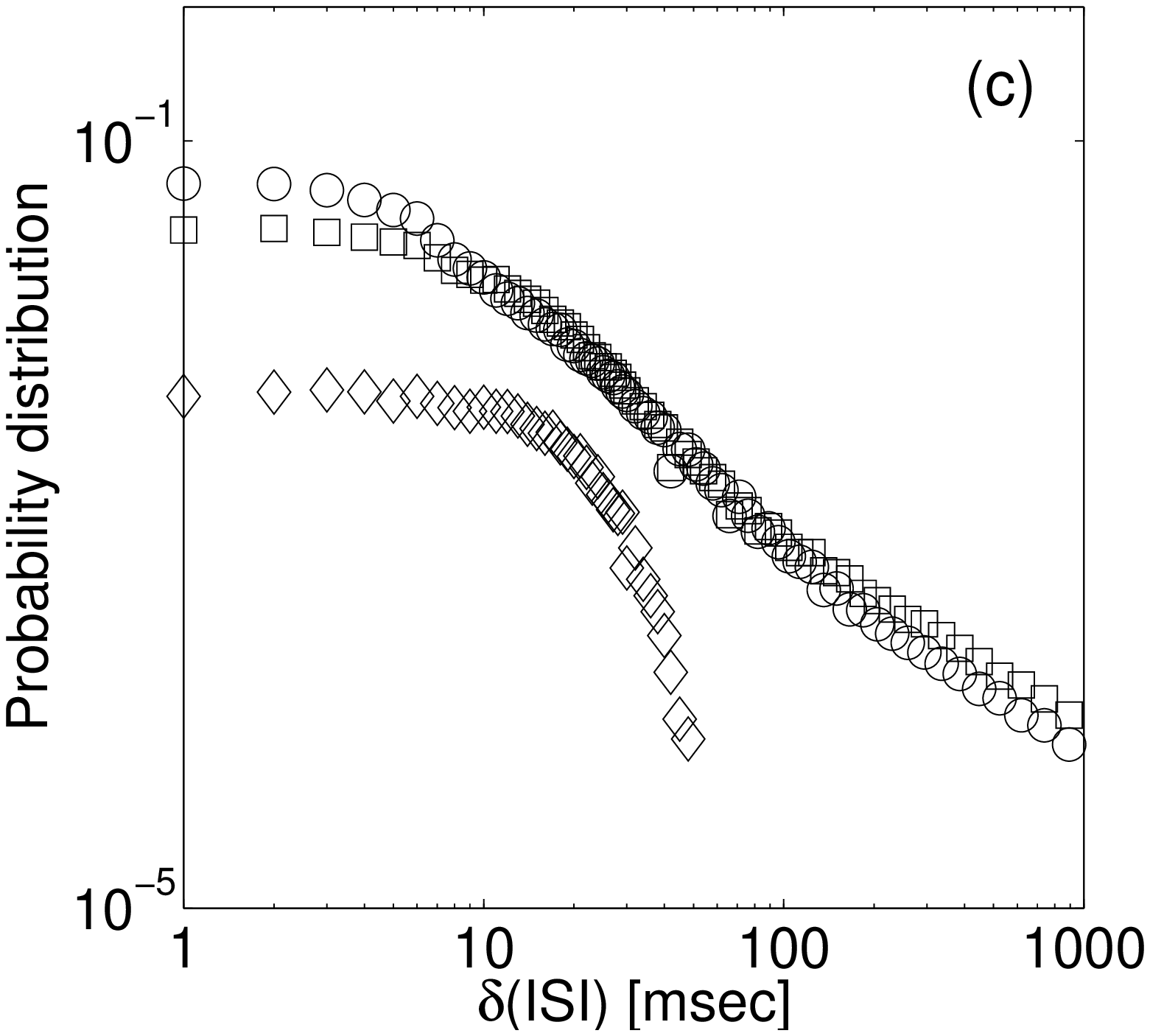}}}
\caption[]{ The activity of a model neuron containing the
self-synapse (autapse), as modelled by the "classical"
Tsodyks-Uziel-Markram model of synaptic transmission. In this
case, it is possible to recover some of the features of cortical
rapidly-firing neurons, namely the relatively high-frequency
persistent activity. However, the resulting time-series of action
potentials for such a model neuron, shown in (a), is almost
periodic. Due to the self-synapse, a periodic series of spikes
results in the periodic pattern for the post-synaptic current
(shown in (b)), which closes the self-consistency loop by causing
a model neuron to generate a periodic time-series of spikes.
Further difference between the model neuron and between cortical
rapidly-firing neurons is seen upon comparing the corresponding
distributions of ISI increments, plotted on double-logarithmic
scale. These distributions, shown in (c), disclose that, contrary
to the cortical rapidly-firing neurons, the increments
distribution for the model neuron with TUM autapse (diamonds) is
Gaussian (seen as a "stretched" parabola on double-log scale),
pointing at the existence of characteristic time-scale. On the
other hand, distributions for cortical neurons (squares and
circles) decay algebraically and are much broader. The
distribution of the model neuron has been vertically shifted, for
clarity of comparison. } \label{fig-Fig3}
\end{figure}

We focus on an excitatory neuron modelled with Morris-Lecar
dynamics, as described in Model section. We add some external bias
current so as to place the neuron in a state slightly beyond the
saddle-node bifurcation, to where it would spontaneously oscillate
at a very low frequency in the absence of any synaptic input. We
then assume that this neuron has a self-synapse (autapse). An
excitatory self-synapse clearly has the possibility of causing a
much higher spiking rate than would otherwise be the case; this
behavior without any astrocyte influence is shown in Fig.
\ref{fig-Fig3}. The existence of autaptic neurons was originally
demonstrated in cultured
networks~\cite{Segal91,Bekkers91,Segal94}, but has been detected
in intact neocortex as well~\cite{Lubke96}. Importantly, these can
be either inhibitory or excitatory. There has been some
speculation regarding the role of autapses in
memory~\cite{Seung00}, but this is not our concern here.

\begin{figure}
\centerline{
\resizebox{0.45\textwidth}{!}{\includegraphics{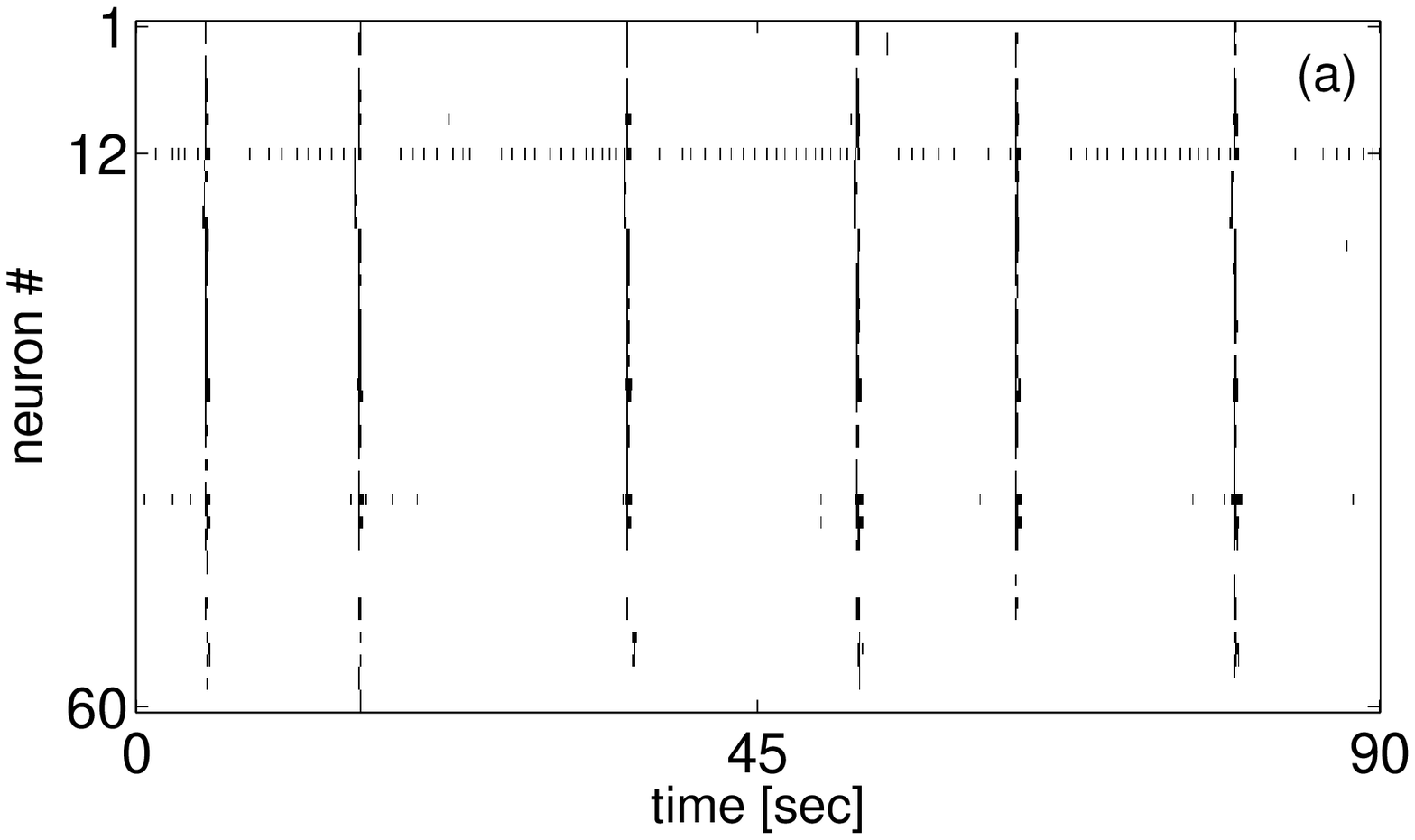}}}
\vspace{1cm}
\centerline{\resizebox{0.350\textwidth}{!}{\includegraphics{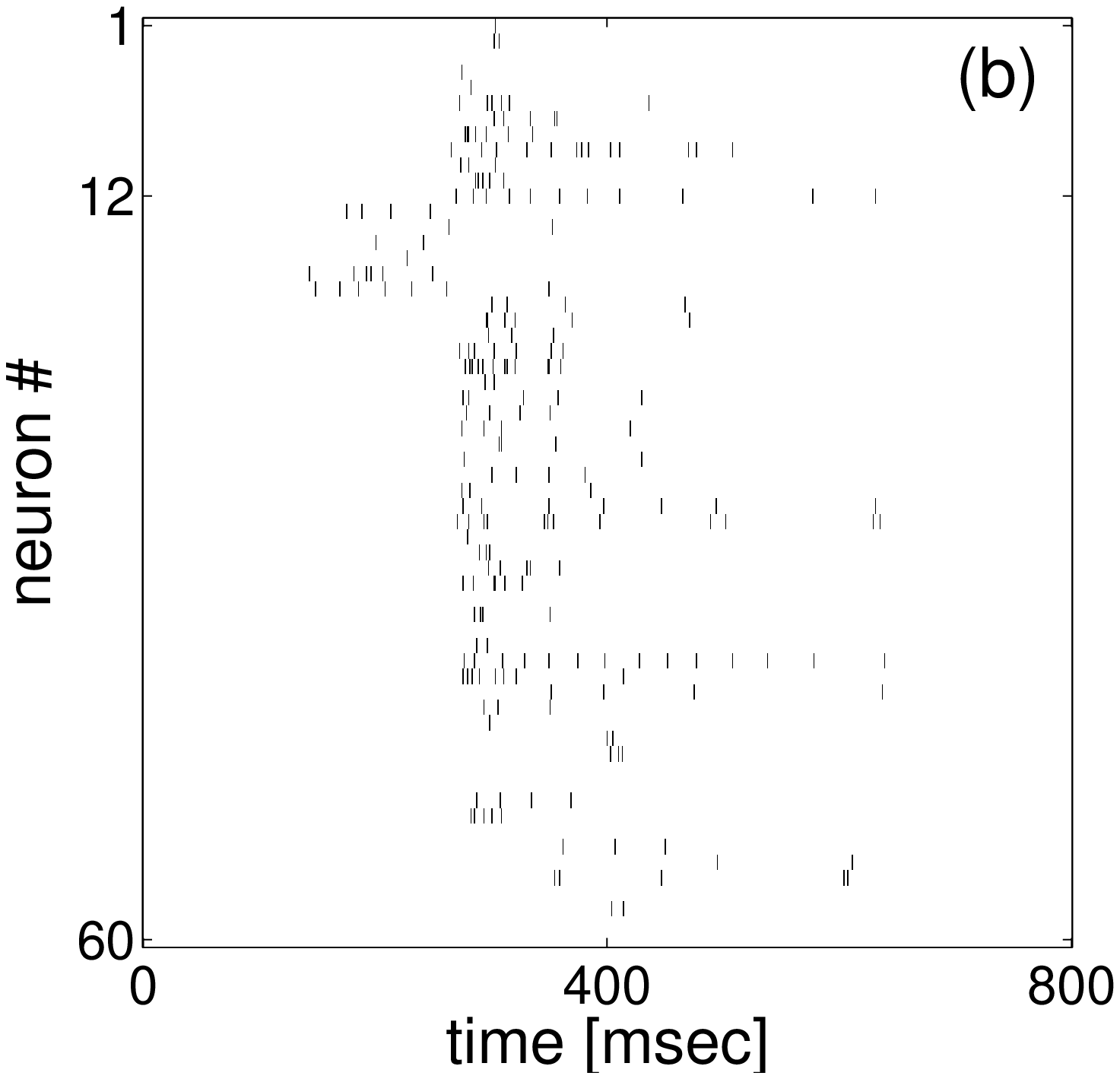}}
\hspace{0.5cm}\resizebox{0.350\textwidth}{!}{\includegraphics{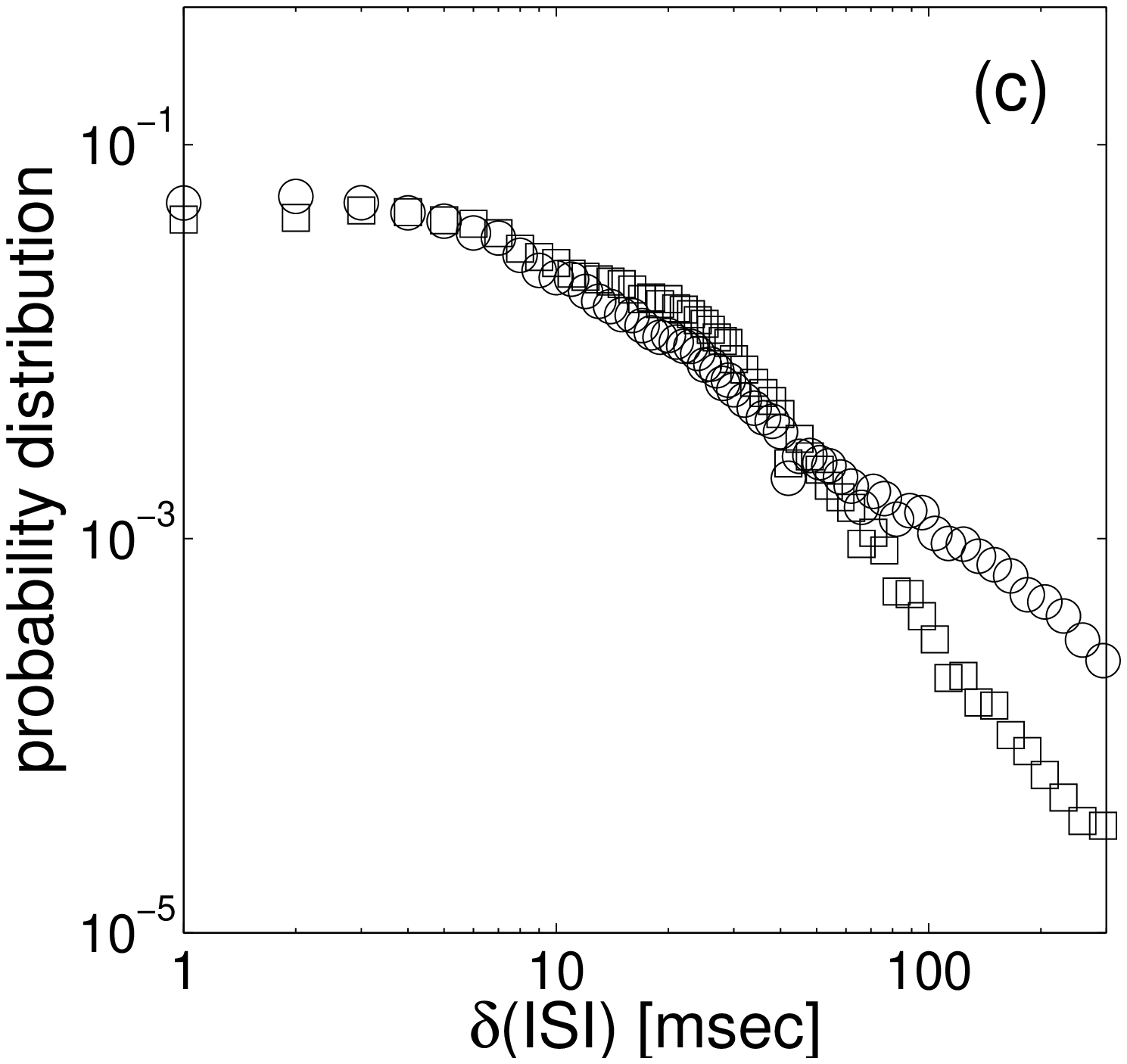}}}
\caption[]{ Electrical activity of \textit{in-vitro} cortical
networks. These cultured networks are spontaneously formed from a
dissociated mixture of cortical neurons and glial cells drawn from
one-day-old Charles River rats. The cells are homogeneously spread
over a lithographically specified area of Poly-D-Lysine for
attachment to the recording electrodes. The activity of a network
is marked by formation of synchronized bursting events $(SBEs)$,
short $(\sim100-400msec)$ periods of time during which most of the
recorded neurons are active. a) A raster plot of recorded
activity, showing a sample of few $SBEs$. The time axis is divided
into $10^{-1}s$ bins. Each row is a binary bar-code representation
of the activity of an individual neuron, i.e. the bars mark
detection of spikes. Note that, while majority of the recorded
neurons are firing rapidly mostly during $SBEs$, there are some
neurons that are marked by persistent intense activity (for
example neuron no.12). This property supports the notion that the
activity of these neurons is autonomous and hence self-amplified.
b) A zoomed view of a sample synchronized bursting event. Note
that each neuron has its own pattern of activity during the $SBE$.
To access the differences in activity between ordinary neurons and
neurons that show intense firing between the $SBEs$, for each
neuron we constructed the series of increments of inter-spike
intervals (ISI), defined as $\delta(i)=ISI(i+1)-ISI(i), i\geq 1$.
The distributions of $\delta(i)$, shown in (c), disclose that the
dynamics of ordinary neurons (squares) is similar to the dynamics
of rapidly firing neurons (circles), up to the time-scale of
100msec, corresponding to the width of a typical $SBE$. Note that
since \textit{increments} of inter-spike-intervals are analyzed,
the increased rate of neurons firing does not necessarily affect
the shape of the distribution. Yet, above the characteristic time
of 100msec, the distributions diverge, possibly indicating the
existence of additional mechanisms governing the activity of
rapidly-firing neurons on a longer time-scale. Note that for
normal neurons there is another peak at typical inter-burst
intervals ($>$ seconds), not shown here.} \label{fig-Fig4}
\end{figure}

Are such neurons observed experimentally? In Fig. \ref{fig-Fig4}
we show a typical raster plot recorded from cultured neural
network grown from a dissociated mixture of glial and neuronal
cortical cells taken from one day old Charles River rats (see
Experimental Observations). The spontaneous activity of the
network is marked by synchronized bursting events (SBEs)  - short
(several 100s of ms) periods during which most of the recorded
neurons show relatively rapid firing separated by long (order of
seconds) time intervals of sporadic neuronal firing of most of the
neurons~\cite{Segev01}. Only small fractions of special neurons
(termed spiker neurons) exhibit rapid firing also during inter
ÐSBEs intervals. These spiker neurons also exhibit much higher
firing rates during the SBEs. But the behavior of these rapid
firing neurons do not match that expected of the simple autaptic
oscillator. The major differences, as illustrated by comparing
Figs. \ref{fig-Fig3} and \ref{fig-Fig4} are 1. the existence of
long inter-spike-intervals for the spikers, marked by a long tail
(Levy) distribution of the increments of the inter-spike
Ðintervals. 2. The beating or burst-like rate modulation in the
temporal ordering of the spike train.

\begin{figure}
\centerline{\resizebox{0.4\textwidth}{!}{\includegraphics{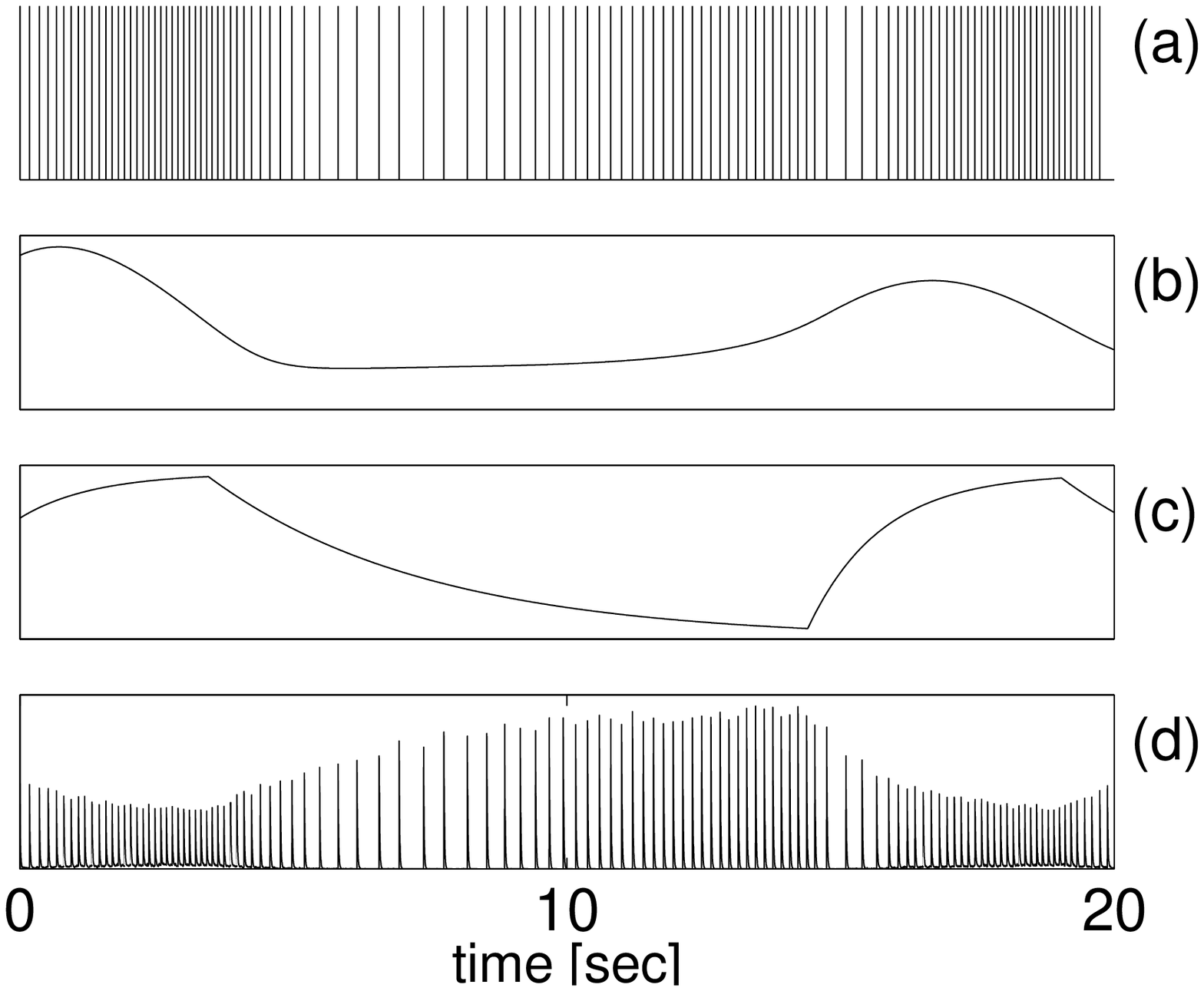}}}
\vspace{0.2cm}
\centerline{\resizebox{0.4\textwidth}{!}{\includegraphics{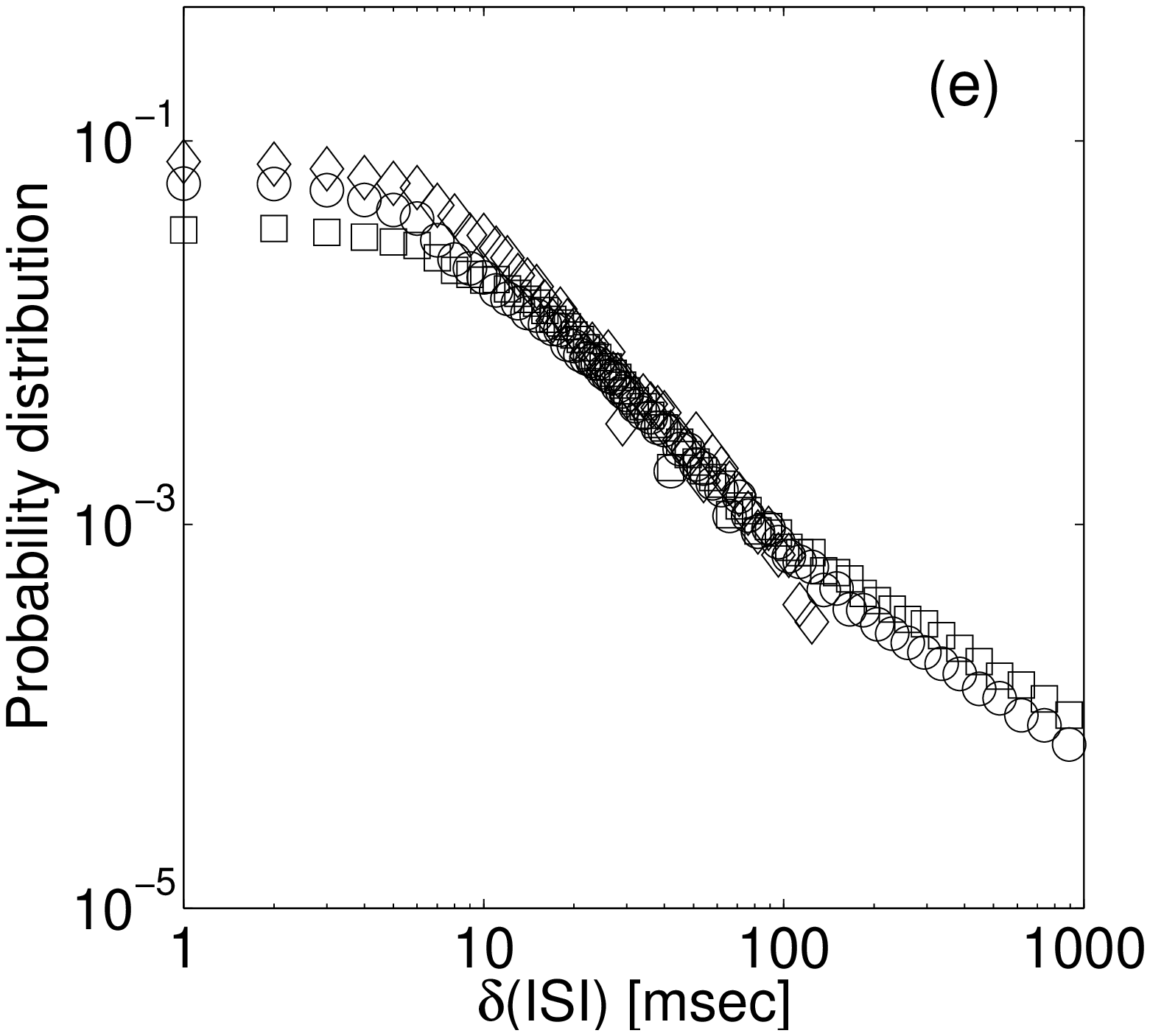}}}
\caption[]{ The activity of a model neuron containing a glia-gated
autapse. The equations of synaptic transmission for this case have
been modified to take into account the influence of
synaptically-associated astrocyte, as explained in text. The
resulting spike time-series, shown in (a), deviates from
periodicity due to the slow modulation of the synapse by the
adjacent astrocyte. The relatively intense activity at the
pre-synaptic terminal activates astrocyte receptors, which in turn
leads to the production of $IP_{3}$ and subsequent oscillations of
free astrocyte $Ca^{2+}$ concentration. The period of these
oscillations, shown in (b), is much larger than the characteristic
time between spikes arriving at the pre-synaptic terminal. Because
$Ca^{2+}$ dynamics is oscillatory, so also will be the dynamics of
the gating function $f$, as is seen from (c), and period of
oscillations for $f$ will follow the period of $Ca^{2+}$
oscillations. The periodic behavior of $f$ leads to slow periodic
modulation of PSC pattern (shown in (d)), which closes the
self-consistency loop by causing a neuron to fire in a burst-like
manner. Additional information is obtained after comparison of
distributions for ISI increments, shown in (e). Contrary to
results for the model neuron with a simple autapse (see figure
\ref{fig-Fig4}c), the distribution for a glia-gated autaptic model
neuron (diamonds) now closely follows the distributions of two
sample recorded cortical rapidly-firing neurons (squares and
circles), up to the characteristic time of $\sim 100msec$, which
corresponds to the width of a typical SBE. The heavy tails of the
recorded distributions above this characteristic time indicate
that network mechanisms are involved in shaping the form of the
distribution on longer time-scales.} \label{fig-Fig5}
\end{figure}

Motivated by the above and the glial gate-keeping effect studied
earlier, we proceed to test if an autaptic oscillator with a
glia-regulated self-synapse will bring the model into better
agreement. In Fig. \ref{fig-Fig5} we show that indeed the activity
of such a modified model does show the additional modulation. The
basic mechanism results from the fact that after a period of rapid
firing of the neuron, the astrocyte intracellular $Ca^{2+}$
concentration (shown in Fig. \ref{fig-Fig5}b) exceeds the critical
threshold for time-delayed attenuation. This then stops the
activity and gives rise to large inter-spike intervals. The
distributions shown in Fig. \ref{fig-Fig5} are a much better match
to experimental data for time intervals up to 100 msec.

\section*{\Large {Robustness tests}}
{\bf The stochastic Li-Rinzel model:} One of the implicit
assumptions of our model for astrocyte-synapse interaction is
related to the deterministic nature of astrocyte calcium release.
It is assumed that in the absence of any $IP_{3}$ signals from the
associated synapses, the astrocyte will stay "silent", in the
sense that there will be no spontaneous $Ca^{2+}$ events. However,
it should be kept in mind that the equations for the calcium
channel dynamics used in the context of Li-Rinzel model, in fact
describe the collective behavior of large number of channels. In
reality, experimental evidence indicates that the calcium release
channels in astrocytes are spatially organized in small clusters
of 20-50 channels - the so-called "micro-domains". These
micro-domains were found to contain small membrane leaflets (of
$O(10nm)$ thick), wrapping around the synapses and potentially
being able to synchronize ensembles of synapses. This finding
calls for a new view of astrocytes as cells with multiple
functional and structural compartments.

The micro-domains (within the same astrocyte) have been observed
to generate the spontaneous $Ca^{2+}$ signals. As the passage of
the calcium ions through a single channel is subject to
fluctuations, for small clusters of channels the stochastic
aspects can become important. Inclusion of stochastic effects can
explain the generation of calcium puffs - fast localized
elevations of calcium concentration. Hence, it is important to
test the possible effect of stochastic calcium events on the
model's behavior.

We achieve this goal by replacing the deterministic Li-Rinzel
model with its stochastic version, obtained using Langevin
approximation, as has been recently described by Shuai and Jung
~\cite{ShuaiLNP}. With the Langevin approach, the equation for the
fraction of open calcium channels is modified, and takes the
following form:

\begin{equation}
\frac{dq}{dt}=\alpha_{q}(1-q)-\beta_{q}q+\xi(t)
\end{equation}
in which the stochastic term, $\xi(t)$, has the following
properties:
\begin{equation}
\langle\xi(t)\rangle=0
\end{equation}
\begin{equation}
\langle\xi(t)\xi(t^{'})\rangle=\frac{\alpha_{q}(1-q)+\beta_{q}q}{N}\delta(t-t^{'})
\end{equation}

\begin{figure}
\centerline{\resizebox{0.6\textwidth}{!}{\includegraphics{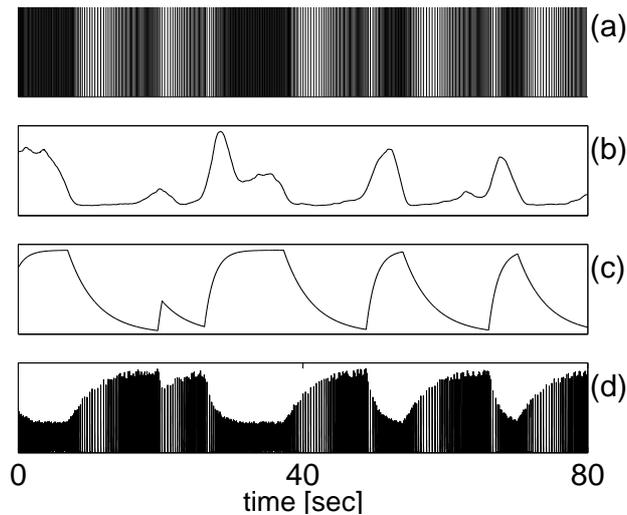}}}
\caption[]{ The dynamical behavior of an astrocyte-gated model
autaptic neuron, including the stochastic release of calcium from
ER of astrocyte. Shown are the results of the simulation when
calcium release from intra-cellular stores is mediated by a
cluster of N=10 channels. The generic form of the spike
time-series (shown in (a)) does not differ from those obtained for
the deterministic model. Namely, even for the stochastic model the
neuron is still firing in a burst-like manner. Although the
temporal profile of astrocyte calcium (b) is irregular, the
resulting dynamics of the gating function (c) is relatively
smooth, stemming from the choice of the gating function dynamics
(being an integration over the calcium profile). As a result, the
PSC profile (shown in d) does not differ much from the
corresponding PSC profile obtained for the deterministic model.}
\label{fig-Fig6}
\end{figure}

In the limit of very large cluster size, $N\rightarrow\infty$, and
the effect of stochastic $Ca^{2+}$ release is not significant. On
the contrary, the dynamics of calcium release are greatly modified
for small cluster sizes. A typical spike time-series of glia-gated
autaptic neuron, obtained for the cluster size of $N=10$ channels,
is shown Fig. \ref{fig-Fig6}a. Note that, while there appear
considerable fluctuations in concentration of astrocyte calcium
(Fig. \ref{fig-Fig6}b), the dynamics of the gating function (Fig.
\ref{fig-Fig6}c) is less irregular. This follows because our
choice of the gating function corresponds to the integration of
calcium events. We have also checked that the distribution of inter-spike-intervals are practically unchanged (data not shown). All told, our results indicate including the stochastic nature of the release of calcium from astrocyte ER does not affect the dynamics of our model
autaptic neuron in any significant way.

{\bf The correlation time of the gating function:} Another
assumption made in our model concerns the dynamics of the gating
function. We have assumed the simple first-order differential
equation for the dynamics of our phenomenological gating function,
and have selected time-scales that are believed to be consistent
with the influence of astrocytes on synaptic terminals. However,
because the exact nature of the underlying processes (and
corresponding time-scales) is unknown, it is important to test the
robustness of the model to variations in the gating function
dynamics.

\begin{figure}
\centerline{\resizebox{0.65\textwidth}{!}{\includegraphics{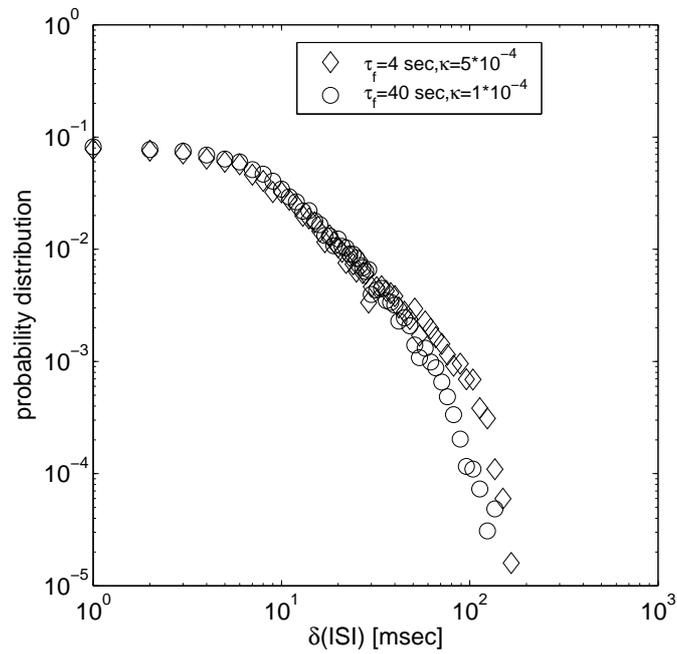}}}
\caption[]{ Distributions of inter-spike-interval increments for
the model of an astrocyte-gated autaptic neuron with slow dynamics
of the gating function, as compared with the corresponding
distribution for the deterministic Li-Rinzel model. Due to the
slow dynamics of the gating function, the transitions between
different phases of bursting are blurred, resulting in a weaker
tail for the distribution of inter-spike interval increments.}
\label{fig-Fig9}
\end{figure}

To do that, we altered the baseline dynamics of the gating
function to have a slower characteristic decay time and a slower
rate of accumulation; for example, we can set  $\tau_{f}=40 sec$
and $\kappa=0.1 sec^{-1}$.  Simulations show that the only effect
is a slight blurring of the transition between different phases of
the bursting, as would be expected. This can best be detected by
looking at the distribution of inter-spike-interval increments,
for the case of slow gating dynamics. The distribution, shown in
Fig. \ref{fig-Fig9}, has weaker tail as compared to the
distribution obtained for the faster gating dynamics. This result
follows because for a slower gating, the modulation of the
post-synaptic current is weaker. Hence, the transitions from
intense firing to low-frequency spiking are less abrupt, resulting
in a relatively low proportion of large increments.It is worth
remembering that large increments of inter-spike intervals reflect
sudden changes in dynamics which are eliminated by the blurring.
Clearly, the model with fast gating does a better job in fitting
the spiker data.

{\bf The time-dependent background current:} All of the main
results were obtained under the assumption of constant background
current feeding into neuronal soma, such that when acting alone,
this current forces the model neuron to fire at some very low
frequency. One may justly argue that there is no such thing as
constant current. Indeed, if a background current has to do with
the biological reality, then it should possess some dynamics. For
example, a better match would be to imagine the background current
to be associated with the activity in adjacent astrocytes (see
e.g. \cite{Angulo04}).

\begin{figure}
\centerline{\resizebox{0.6\textwidth}{!}{\includegraphics{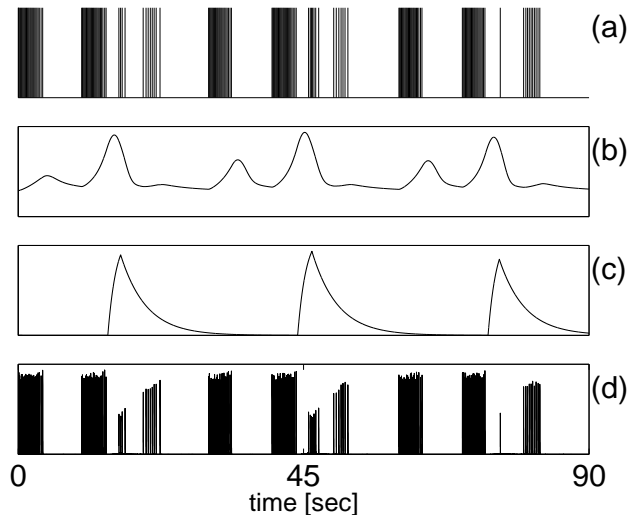}}}
\caption[]{ The dynamical behavior of an astrocyte-gated model
autaptic neuron with slowly oscillating background current. Shown
are the results of the simulation when $I_{base}\propto
sin(\frac{2\pi}{T}t)$, $T=10 sec$. The mean level of $I_{base}$ is
set so as to put a neuron in the quiescent phase for half a
period. The resulting spike time-series (shown in a) disclose the
burst-like firing of a neuron, with the super-imposed oscillatory
dynamics of a background current. The variations in the
concentration of astrocyte calcium (b) are much more temporally
localized, and so is the resulting dynamics of the gating function
(shown in c). Consequently, the PSC profile (d) strongly reflects
the burst-like synaptic transmission efficacy, thus forcing the
neuron to fire in a burst-like manner and closing the
self-consistency loop.} \label{fig-Fig10}
\end{figure}

To test this, we simulated glia-gated autaptic neuron subject to
slowly oscillating $(T=10 sec)$ background current. For this case,
we found that the behavior of a model is generically the same.
Yet, now the transitions between the bursting phases are sharper
(see Fig. \ref{fig-Fig10}a). This, in turn, leads to the sharper
modulation of post-synaptic currents (shown in Fig.
\ref{fig-Fig10}d). We can confirm this by noting that the distribution of
inter-spike interval increments has a
slightly heavier tail, as compared to the distribution obtained
for the case of constant background current (data not shown). On the other hand,
replacing the constant current with the oscillating one,
introduces a typical frequency, not seen in the actual spiker data. This artificial problem
will presumably disappear when the background current is
determined self-consistently as part of the overall network
activity. Similarly, the key to extending the increments
distributions to longer time scales seems to be getting the
network feedback to the spikers to regulate the inter-burst timing
which at the moment is too regular. This will be presented in a
future publication.

\section*{\Large {Discussion}}

In this paper, we have proposed that the regulation of synaptic
transmission by astrocytic calcium dynamics is a critical new
component of neural circuitry. We have used existing biophysical
experiments to construct a coupled synapse-astrocyte model to
illustrate this regulation and to explore its consequences for an
autaptic oscillator, arguably the most elementary neural circuit.
Our results can be compared to data taken from cultured neuron
networks. This comparison reveals that the glial ``gate-keeping"
effect appears to be necessary for an understanding of the
inter-spike interval distribution of observed rapidly-firing
``spiker" neurons, for time scales up to about 100msec.

Of course, many aspects of our modelling are quite simplified as
compared to the underlying biophysics. We have investigated the
sensitivity of our results to the modification of some of the
parameters of our model as well as the addition of more complex
dynamics for the various parts of our system. Our results with
regard to the inter-spike interval are exceedingly robust.

This work should be viewed as a step towards understanding the
full dynamical consequences brought about by the strong reciprocal
couplings between synapses and the glial processes that envelop
them. We have focused on the fact that astrocytic emissions shut
down synaptic transmission when the activity becomes too high.
This mechanism appears to be a necessary part of the regulation of
spiker activity; without it, spikers would fire too often, too
regularly. Related work by Nadkarni and Jung (private
communication) focuses on a different aspect, that of increased
fidelity of synaptic release (for otherwise highly stochastic
synapses) due to glia-mediated increases in pre-synaptic calcium
levels. As our working assumption is that the spikers are most
likely to be neurons with ``faithful" autapses, this effect does
not play a role in our attempt to compare to the experimental
data. It will of course be necessary to combine these two
different pieces to obtain a more complete picture.

As stressed here, the application to spikers is just one way in
which our new synaptic dynamics may alter our thinking about
neural circuits. This particular application is appealing and
informative but must at the moment be considered an untested
hypothesis. Future experimental work must test the assumption that
spikers have significant excitatory autaptic coupling, that
pharmacological blockage of the synaptic current reverts their
firing to low frequency almost periodic patterns, and that cutting
the feedback loop with the enveloping astrocyte eliminates the
heavy-tail increment distribution. Work towards achieving these
tests is ongoing.

In the experimental system, a purported autaptic neuron is a part
of active network  and would therefore receive input currents from
the other neurons in the network. This more complex input would
clearly alter the very-long-time inter-spike interval
distribution, especially given the existence of a new {\em
inter-burst} timescale in the problem. Similarly, the current
approach of adding a constant background current to the neuron is
not realistic; the actual background current, due to such
processes as glial-generated currents in the cell soma, would
again alter the long-time distribution. Preliminary tests have
shown that these effects could extend the range of agreement
between autaptic oscillator statistics and experimental
measurements.

Just as the network provides additional input for the spiker, the
spiker provides part of the stimulation that leads to the bursting
dynamics. Future work will endeavor to create a fully
self-consistent network model to explore the overall activity
patterns of this system. One issue that needs investigation
concerns the role that glia might have in coordinating the action
of neighboring synapses. It is well-known that a single astrocytic
process might contact thousands of synapses; if the calcium
excitation spreads from being a local increase in a specific
terminus to being a more widespread phenomenon within the glial
cell body, neighboring synapses can become dynamically coupled.
The role of this extra complexity in shaping the burst structure
and time sequence is as yet unknown.

\section*{{\Large Acknowledgements}}
The authors would like to thank Gerald M. Edelman for insightful
conversation about the possible role of glial cells. Eugene
Izhikevich, Peter Jung, Suhita Nadkarni, Mark Shein, Nadav
Raichman and Itay Baruchi are acknowledged for useful comments and
for the critical reading of an earlier version of this manuscript.
Vladislav Volman thanks the Center for Theoretical Biological
Physics for hospitality. This work has been supported in part by
the NSF-sponsored Center for Theoretical Biological Physics (grant
numbers PHY-0216576 and PHY-0225630), by Maguy-Glass Chair in
Physics of Complex Systems.

\section*{{\Large Parameters used in simulations}}
\begin{tabular}{|c|c|c|c|c|c|}
   \hline
  $P_{0}$ & $0.5$ & $\eta_{mean}$ & $1.2\cdot10^{-3}\mu Acm^{-2}$ & $\sigma$ & $0.1$ \\
   \hline
  $\tau_{ca}$ & $4sec$ & $\kappa$ & $0.5sec^{-1}$ & $\tau_{IP_{3}}$ & $7sec$ \\
   \hline
  $r_{IP_{3}}$ & $7.2 mMsec^{-1}$ & $IP_{3}^{*}$ & $0.16 \mu M$ & $c_{1}$ & $0.185$  \\
   \hline
  $v_{1}$ & $6sec^{-1}$ & $v_{2}$ & $0.11sec^{-1}$ & $v_{3}$ & $0.9\mu Msec^{-1}$ \\
   \hline
  $k_{3}$ & $0.1\mu M$ & $d_{1}$ & $0.13\mu M$ & $d_{2}$ & $1.049\mu M$ \\
   \hline
  $d_{3}$ & $0.9434\mu M$ & $d_{5}$ & $0.08234\mu M$ & $a_{2}$ & $0.2\mu M^{-1}sec^{-1}$ \\
   \hline
  $c_{0}$ & $2.0\mu M$ & $g_{Ca}$ & $1.1 mS cm^{-2}$ & $g_{K}$ & $2.0 mS cm^{-2}$ \\
   \hline
  $g_{L}$ & $0.5 mS cm^{-2}$ & $V_{Ca}$ & $100 mV$ & $V_{K}$ & $-70 mV$ \\
   \hline
  $V_{L}$ & $-35 mV$ & $V_{1}$ & $-1 mV$ & $V_{2}$ & $15 mV$ \\
   \hline
  $V_{3}$ & $10 mV$ & $V_{4}$ & $14.5 mV$ & $\phi$ & $0.3$ \\
   \hline
  $I_{base}$ & $0.34 \mu A cm^{-2}$ & $\tau_{d}$ & $10msec$ & $\tau_{rec}$ & $100msec$ \\
   \hline
  $u$ & $0.1$ & $A$ & $10 \mu Acm^{-2}$ & \\
   \hline
\end{tabular}


\bibliography{spiker-bib}

\newpage


\newpage

\newpage

\newpage

\newpage

\newpage

\newpage

\newpage

\end{document}